\documentclass[conference]{IEEEtran}
\IEEEoverridecommandlockouts
\usepackage{cite}
\usepackage{amsmath,amssymb,amsfonts}
\usepackage{algorithmic}
\usepackage{graphicx}
\usepackage{textcomp}
\usepackage{epsfig}
\usepackage{epstopdf}
\usepackage{xcolor}
\def\BibTeX{{\rm B\kern-.05em{\sc i\kern-.025em b}\kern-.08em
    T\kern-.1667em\lower.7ex\hbox{E}\kern-.125emX}}

\def\be{\begin{equation}}
\def\ee{\end{equation}}

\def\mih{\mathbf{h}}

\def\mix{\mathbf{x}}

\def\mi0{\mathbf{0}}

\begin{document}

\title{ Resource Allocation in Laser-based Optical Wireless Cellular Networks
{\footnotesize \textsuperscript{}}
 
 \thanks{This work  has been supported in part by the Engineering and Physical Sciences Research Council (EPSRC), in part by the INTERNET project under Grant EP/H040536/1, and in part by the STAR project under Grant EP/K016873/1 and in part by the TOWS project under Grant EP/S016570/1. All data are provided in full in the results section of this paper.}
}

\author{\IEEEauthorblockN{Ahmad Adnan Qidan$ ^1 $, M\'aximo Morales-C\'espedes$ ^2 $, Taisir El-Gorashi$ ^1 $, Jaafar M. H. Elmirghani$ ^1 $}
\IEEEauthorblockA{$ ^1 $School of Electronic and Electrical Engineering, University of Leeds, LS2 9JT, United Kingdom 
\\$ ^2 $Department of Signal Theory and Communications, Universidad Carlos III de Madrid, Legan\'es, Spain
\\Email: a.a.qidan@leeds.ac.uk, maximo@tsc.uc3m.es, t.e.h.elgorashi@leeds.ac.uk ,j.m.h.elmirghani@leeds.ac.uk}


}

\maketitle

\maketitle

\begin{abstract}
Optical wireless communication provides data transmission at high speeds
which can satisfy the increasing demands 
 for connecting a massive number of devices to the Internet. In this paper, vertical-cavity surface-emitting(VCSEL) lasers are used as transmitters due to their high modulation speed and energy efficiency. However, a high number of VCSEL lasers is  required to ensure  coverage where each laser source illuminates a confined area. Therefore, multiple users are classified into different  sets according to their connectivity. Given this point,  a transmission scheme that uses blind interference alignment (BIA) is implemented to manage the interference in the laser-based network. In addition, an optimization problem is formulated to maximize the utility sum rate taking  into consideration the classification of the users. To solve this problem, a decentralized algorithm is  proposed where the main problem is divided into sub-problems, each can be solved independently avoiding complexity. The results demonstrate the optimality of the decentralized algorithm where a sub-optimal solution is provided. Finally, it is shown that BIA can provide high performance in laser-based networks compared with  zero forcing (ZF) transmit precoding scheme. 
\end{abstract}

\IEEEpeerreviewmaketitle

\section{Introduction} 
In recent years, the demands for high speed cellular networks has increased significantly  due to the tremendous usage of the Internet. The current cellular networks using radio frequency (RF) are already overloaded, and the optical band is being considered for the next generation of wireless communications in 6G. Transmission based on using light emitting diodes (LEDs) faces various limitations such as the limited bandwidth of these light sources. In the meantime, research is underway towards using laser sources, namely Vertical-Cavity Surface-Emitting  (VCSEL)  lasers with transmitted power that ensures human eye safety \cite{Liu:19,9217158}. 

In this context, VCSELs have a small coverage area which might be limited to a few centimeters. Therefore, multiple optical access points (APs) must be deployed in an indoor environment to ensure the coverage. Given this point, the performance of optical wireless networks can be subject to high inter-cell interference (ICI) resulting in poor signal to noise ration (SNR). Besides, considering multi-user multiple-input multiple-output (MU-MIMO) optical systems, interference management among the users must be addressed. In \cite{8636954,9064520,1990icc}, blind interference Alignment (BIA) is implemented for optical networks based on various network topologies to manage the interference without channel state information (CSI). It is shown that BIA  schemes  characterized by their positive precoding matrices ensure the non-negativity of transmitted signal. In addition, cooperation among optical APs  is not required for BIA schemes while maximizing the multiplexing gain without exchanging the  information of users among transmitters. As a result, BIA schemes achieve greater sum-rate than other schemes such as zero forcing (ZF) and maximum rate combining (MRC) in optical scenarios \cite{8636954,9064520,1990icc}.

In general, the performance of such cellular networks can be enhanced by addressing the resource management problem among users. Specifically,  several optimization problems have been formulated in  optical cellular systems with the aim of maximizing the utility-based sum rate of users \cite{6933944,9064520,7511380,JLT1111}. In \cite{6933944,9064520}, an optimization problem is formulated  to allocate the resources, thus  maximizing the overall rates of heterogeneous optical/RF networks. In \cite{7511380}, power allocation schemes  are implemented  to maximize the user rate while providing uniform illumination. Furthermore, in \cite{JLT1111},  an optimization problem is formulated 
in an optical scenario composed of multiple optical APs for resource allocation while satisfying the requirements of multiple users. However, resource allocation is still an open issue that needs further investigation, especially in laser-based scenarios where each laser covers a small area.  

In this work, BIA is implemented to align the interference among users served by multiple VCSEL lasers considering the resource allocation problem. First, the users are divided into two sets denoted as full connectivity and partial connectivity groups, which are defined by the connectivity of those users. 
After that, an optimization problem is formulated to allocate the resources and find the optimal user assignment considering the two different sets of users. This problem is known as a mixed integer non-linear programming (MINLP) problem, which is difficult and not easy to solve. Given this point, a decentralized algorithm is proposed dividing the  main optimization problem into two sub-problems, each can be solved independently using the Lagrangian function. Simulation  results demonstrate that high performance is achieved using BIA to align the interference among users. Finally, the  decentralized algorithm provides solution significantly close to the optimal with low computational complexity. 

\begin{figure}[t]
\begin{center}\hspace*{0cm}
\includegraphics[width=0.65\linewidth]{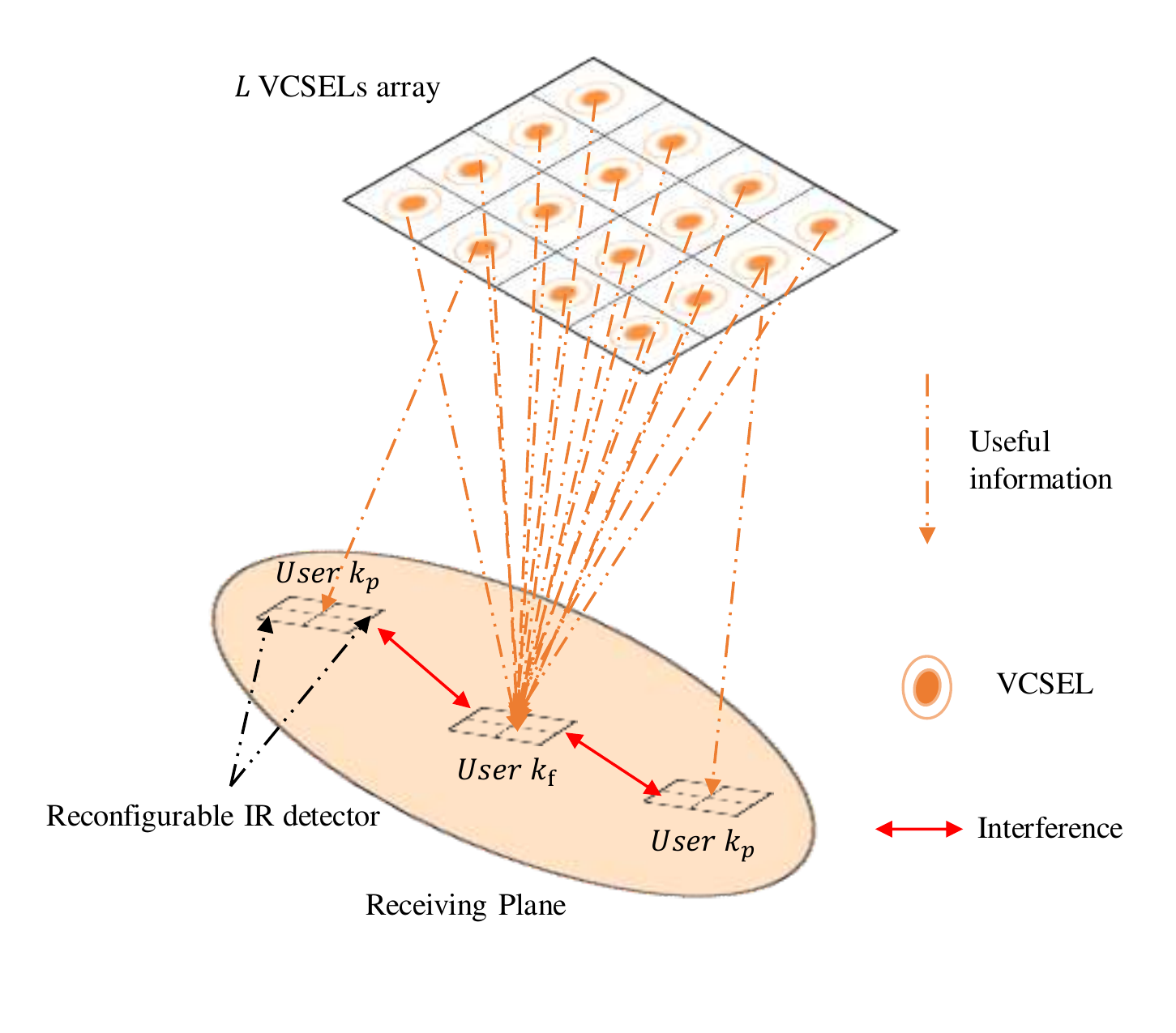}
\end{center}
\vspace{-2mm}
\caption{VCSELs serving multiple users, each user is  equipped with a reconfigurable IR detector.}\label{Figmodel}
\vspace{-2mm}
\end{figure} 

\section{System Model}
\label{sec:system}

The coverage area of each optical AP is limited to a small and confined area. Specifically, we consider  a set of  VCSEL lasers given by $ L $, $ l=\{1, \dots, L\} $, forming an array as shown in Fig. \ref{Figmodel}. Each VCSEL is managed as an optical AP due to the eye safety requirements. This array of VCSELs serves multiple users given  by $ K $, $ k=\{1, \dots, K\} $, distributed randomly on the receiving plane. A reconfigurable optical detector composed of multiple photodiodes given by $ M $, $ m=\{1, \dots, M\} $ , is considered for each user to provide a wide field of view (FoV)\cite{8636954}. For the proposed detector, each photodiode points to a distinct different direction in order to provide linearly independent channel responses.

The users are divided into two sets given by full and partial connectivity composed of  $ K_{f} $ and $ K_{p} $ users, respectively. In particular,  $ K_{f}$ denotes the users allocated in the center of the coverage area receiving useful signals from $ L $ VCSELs, while the users belonging to $ K_{p} $ are connected to the best VCSEL available at their location based on an optimization problem formulated in Section IV. The signal received by a generic user $ k $, $ k \in K $ at time $ n $ connected to $ l $ transmitters is expressed as 


\begin{equation}
y^{[k,l]}[n]=\mih^{[k,l]}(m^{[k,l]}[n])^{T} \mix[n]+  z^{[k,l]}[n],
\end{equation}
where $ \mih^{[k,l]}(m^{[k,l]}[n])^{T} \in \mathbb{R}_+^{l\times 1} $, $ m^{[k,l]}[n] $  is the mode selected by photodiode $ m $ at time slot $ n $, $ \mix $ is the transmitted signal and $ z^{[k,l]} $ is real valued additive white Gaussian noise with zero mean and variance given by 

\begin{equation}
 \sigma^{2}_{z}=\sigma_{\sum}+\mathrm{RIN} \left(\sum^{L}_{l', l'\neq l} ~( h^{[k,l']} ~ \delta_{m}~ P_{t,l'} )^2\right)~B, 
\end{equation}
where $\sigma_{\sum} $  is the sum of the contributions
from both shot noise and thermal noise, and $ \mathrm{RIN} $ is defined as the mean square
of the power fluctuations at a given time divided by the squared
average power, i.e., the relative intensity noise of the VCSEL. Moreover, $ h^{[k,l']} $ is the interference channel received from the neighboring VCSEL laser $ l' $,  $ \delta_{m} $ is the responsivity of photodiode $ m $, $ P_{t,l'} $ is the optical power of VCSEL laser $ l' $ and $ B $ is the single-sided bandwidth of the system.

In this work, the VCSEL lasers are connected to a central unit that controls the resources of the network. Moreover, the interference among users is managed by implementing BIA where the CSI at transmitters is limited to the distribution of users and the channel coherence time. Finally, the reconfigurable detector of each user has the ability to switch among at least $ L=M $ predefined preset modes in order to satisfy the methodology of BIA. 
 
\subsection{Transmitter}
The  optical VCSEL transmitter considered in this work  is  characterized by its  Gaussian beam profile with multiple modes. Basically, the transmitted power of the VCSEL  is based on the beam waist $ W_{0} $, the wavelength $ \lambda $ and the distance $ z $ through which the beam travels. In this sense, the beam radius at photodiode $ m $ of user $ k $ located at distance $d=z$ is given by 

\begin{equation}
W_{d}=W_{0} \left( 1+ \left(\frac{d}{d_{Ra}}\right)^{2}\right)^{1/2},
\end{equation}
where $ d_{Ra} $ is the Rayleigh range given by 
\begin{equation}
d_{Ra}= \frac{\pi W^{2}_{0} n }{ \lambda},
\end{equation}
where $ n $ is the refractive index of the medium, air in this case, which is equal to 1. In general, the intensity of the Gaussian beam is defined as a function of the radial distance $ r $ from the center of the beam spot and the distance $ z $. That is, the spatial distribution of the intensity of VCSEL laser $ l $ over the transverse plane at distance $ d $ is given by 

\begin{equation}
I_{l}(r,d) = \frac{2 P_{t,l}}{\pi W^{2}_{d}}~ \mathrm{exp}\left(-\frac{2 r^{2}}{W^{2}_{d}}\right).
\end{equation}
In particular, considering the  detection area of the reconfigurable detector denoted by $ A_{rec} $, the area of each photodiode is given by $ A_m = \frac{A_{rec}}{M} $, $ m \in M $. Therefore, the power received by the photodiode $ m $  of user $ k $ located at distance $ d $ from  VCSEL  $ l $ is given by 
\begin{equation}
\begin{split}
&P_{m,l}=\\
&\int_{0}^{A_m /2 \pi} I(r,d) 2\pi r dr = P_{t,l}\left[1- \mathrm{exp}\left(- 2 \left(\frac{ A_{m}}{2 \pi W_{d}}\right)^{2}\right)\right],
\end{split}
\end{equation}
assuming the photodiode $ m $ of user $ k $ is located right under transmitter $ l $. At this point, the optical channel between the photodiode $ m $ of user $ k $ and transmitter $ l $ can be given by   
\begin{equation}
h^{[k,l]}(m)=h_{ \mathrm{LoS}}^{[k,l]}(m) + h_{\mathrm{diff}}(f) e^{-j2\pi f \Delta T}, 
\end{equation}
where $h_{ \mathrm{LoS}}^{[k,l]}(m)$ is the channel component given by the Line-of-Sight (LoS) link, $h_{\mathrm{\mathrm{diff}}}$ is the diffuse component and $\Delta T$ is the delay between both channel components. Assuming each user is equipped with a reconfigurable detector, a wide FoV  can be provided, and therefore, the NLoS component can be  neglected for the sake of simplicity.


\begin{figure}[t]
\begin{center}\hspace*{0cm}
\includegraphics[width=0.5\linewidth]{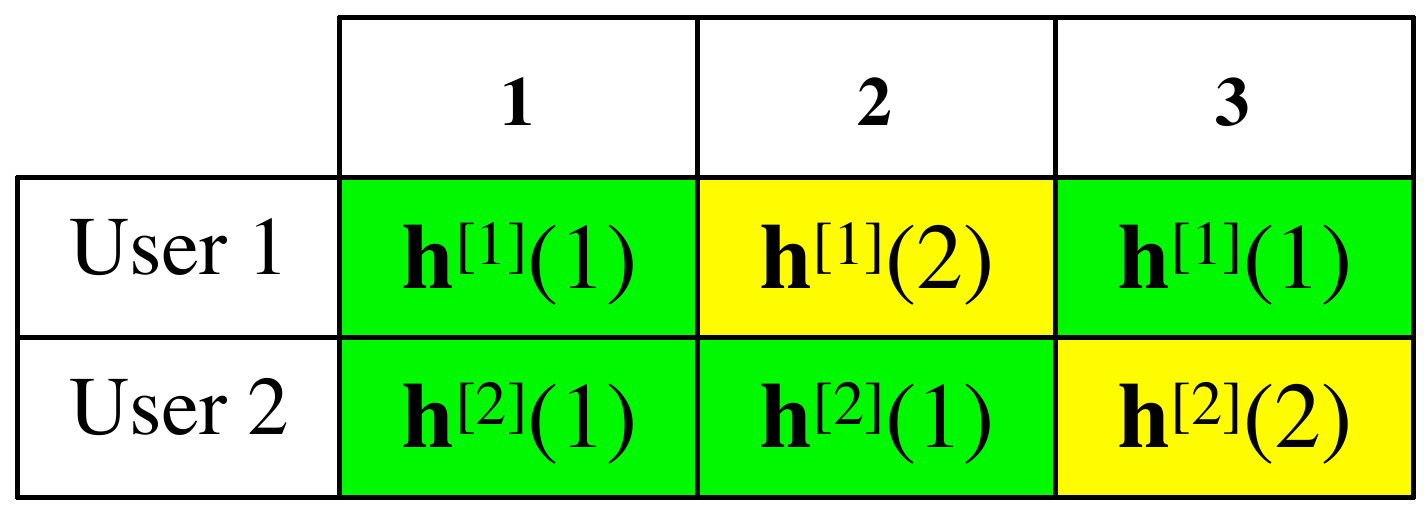}
\end{center}
\vspace{-2mm}
\caption{The supersymbol of BIA for $ L=2 $ and $ K=2 $}\label{bi}
\vspace{-2mm}
\end{figure}
\section{Blind Interference Alignment}
 \label{sec:Blind}
 In this section, the BIA methodology  is defined  to align the interference among users maximizing the multiplexing gain, i.e., DoF. We first describe BIA for a toy example, and then, BIA is applied to our system model deriving the achievable user rate.
 
\subsection{Toy example}
For illustrative purposes, we consider an optical network composed of $ L=2 $ transmitters serving $ K=2 $ users. To align the interference in the absence of CSI at the transmitters, the reconfigurable detector of each user must follow a set of preset modes as shown in Fig. \ref{bi}. Basically, the transmission based on BIA occurs over a supersymbol comprised several symbol extensions, i.e., time slots. The transmitted signal over the supersymbol of this scenario is given by 

\begin{equation}
\mathbf{X} = 
\begin{bmatrix}
\mathbf{x}[1]\\
\mathbf{x}[2]\\
\mathbf{x}[3]\\
\end{bmatrix}
= 
\begin{bmatrix}
\mathbf{I_2}\\
\mathbf{I_2}\\
\mathbf{0_2}\\
\end{bmatrix}
\mathbf{u}_{\ell}^{[1]}
+
 \begin{bmatrix}
\mathbf{I_2}\\
\mathbf{0_2}\\
\mathbf{I_2}\\
\end{bmatrix}
\mathbf{u}_{\ell}^{[2]},
\end{equation} 
where $ \mathbf{x}[n] \in \mathbb{R}_+^{L\times 1} $  is the signal transmitted over the time slot $n  $, $ \mathbf{I_2}  $ and $ \mathbf{0_2} $ are  $ 2 \times 2 $  identity and zero matrices, respectively. Moreover, the first and second time slots compose an alignment block denoted as $ \ell $, over which the  symbol $ \mathbf{u}_{\ell}^{[1]} $ is transmitted to user 1. While the  symbol $ \mathbf{u}_{\ell}^{[2]} $ is transmitted to user 2 over the first and third time slots. Focusing on user 1, the transmitted signal can be expressed as  
\begin{equation}
\begin{bmatrix}y^{[1]}[1]  \\ y^{[1]}[2]  \\ y^{[1]}[3] \end{bmatrix}
= 
\underbrace{
\begin{bmatrix}
\mathbf{h}^{[1]}(1) \\ \mathbf{h}^{[1]}(2) \\ \mathbf{0_{2,1}} 
\end{bmatrix}}_{\mathrm{rank} = 2}
\mathbf{u}_{1}^{[1]}
+ 
\underbrace{
\begin{bmatrix}
\mathbf{h}^{[1]}(1) \\ \mathbf{0_{2,1}} \\ \mathbf{h}^{[1]}(1)  
\end{bmatrix}}_{\mathrm{rank} = 1}
\mathbf{u}_{1}^{[2]}
+
\begin{bmatrix}z^{[1]}[1]  \\ z^{[1]}[2]  \\ z^{[1]}[3] \end{bmatrix}.
\end{equation}
It is worth mentioning that both users are receiving information over the first time slot generating interference. On the other hand, orthogonal transmission is carried out over the second and third time slots giving each user the ability to cancel the interference received over the first time slot. Given this point, the signal received by user 1 after interference subtraction is given by  
\begin{equation}
\label{eq:signalrec}
\begin{bmatrix} y^{[1]}[1]  - y^{[1]}[3] \\ y^{[1]}[2] \end{bmatrix} 
= 
\underbrace{
\begin{bmatrix}
\mathbf{h}^{[1]}(1) \\ \mathbf{h}^{[1]}(2)  
\end{bmatrix}}_{\mathbf{H}^{[1]}}   
\mathbf{u}_{1}^{[1]}
+
\begin{bmatrix} z^{[1]}[1]  - z^{[1]}[3] \\ z^{[1]}[2] \end{bmatrix}, 
\end{equation}
It can be easily seen that the interference is subtracted at the cost of increasing noise. As a consequence, 2 DoF in the symbol $ \mathbf{u}_{\ell}^{[1]} $ can be decoded. Similarly, user 2 can achieve 2 DoF  encoded in the received symbol $ \mathbf{u}_{\ell}^{[1]} $. As a result, 4 DoF can be achieved over a supersymbol that comprises 3 time slots. Therefore, the sum-DoF equals to $ \frac{4}{3} $. Interestingly, 1 DoF can be achieved for the same scenario considering  orthogonal transmission schemes such as TDMA. More details on the implementation of  BIA in the general case are provided in \cite{GWJ11}.   
\subsection{BIA-based user rate of VCSEL}

According to our system model, the users are divided into $ K_f $ and $ K_p $ sets\footnote{Orthogonal transmission is considered to align the interference between $ K_f $ and  $ K_p $ }. In particular, each user belonging to $ K_f $ is connected to all VCSELs due to its location in the center of the coverage area. In this context, the signal received by user $ k_f $, $ k_{f}\in K_{f} $, over an alignment block $ \ell $ that carries $ L  $ DoF from the whole set of  the VCSEL lasers is given by 
\begin{equation}
\label{fl}
\mathbf{y}^{[k_f]} = \mathbf{H}^{[k_f]}\mathbf{u}_{\ell}^{[k_f]} +  \mathbf{z}^{[k_f]},
\end{equation}
where $ \mathbf{H}^{[k_f]} = \begin{bmatrix} \mathbf{h}^{[k_f]}(1) & \dots & \mathbf{h}^{[k_f]}(L)
 \end{bmatrix} \in \mathbb{R}_+^{L\times 1} $ is the channel matrix of user $k_f$ connected to $ L $ VCSELs. It is worth mentioning that the reconfigurable receiver must have at least $ M=L $ photodiodes, giving the ability to each  user to select a signal from $ L $ signals at a time slot, and therefore, $L$ linearly independent channel responses can be contained. In \eqref{fl}, $\mathbf{z}^{[k_f]}$ is the noise after interference subtraction given by the covariance matrix, i.e.,
\begin {equation}
\mathbf{R_{z_f}} = 
\begin{bmatrix} 
K_f\mathbf{I}_{L-1} & \mathbf{0}\\ 
\mathbf{0} & 1\\
\end{bmatrix}.
\end{equation}
For the $ K_p $ set, each user is served by the best link available at its the location. In this sense, the received signal of user $ k_p $, $ k_{p}\in K_{p} $, connected to VCSEL $ l $ is given by   

\begin{equation}
\label{pl}
\mathbf{y}^{[k_{p},l]} = \mathbf{h}^{[k_{p},l]}\mathbf{u}_{\ell}^{[k_{p},l]} +   \sum_{ \substack{l' = 1,  l'\neq l}}^{L}
   \sqrt{\alpha_{l'}^{[k_{p},l]}}\mathbf{h}^{[k_{p},l']}\mathbf{u}_{\ell}^{[k_{p},l']}      +  \mathbf{z}^{[k_{p},l]},
\end{equation}
where $ \alpha_{l'}^{[k_{p},l]} $ is the signal-to-interference (SIR) ratio of user $ k_p $ due to the received interference from VCSEL $ l' $, which is treated as noise. Moreover, $ \mathbf{u}_{\ell}^{[k_{p},l']} $ is the interfering symbol received from the the neighboring VCSEL, and $\mathbf{z}^{[k_{p},l]}$ is the noise after interference subtraction given by the covariance matrix, i.e.,
\begin {equation}
\mathbf{R_{z_p}} = 
\begin{bmatrix} 
K_{p,l}\mathbf{I}_{l-1} & \mathbf{0}\\ 
\mathbf{0} & 1\\
\end{bmatrix}.
\end{equation}
where $ K_{p,l} $ is the total number of users connected to VCSEL $ l $. 

The overall resources allocated to a generic user denoted as $ \nu $, are given by 

\begin{equation}
\label{reseq}
    \nu^{[k]}= 
\begin{cases}
    \  \sum\limits_{l\in L} \nu^{[l,k_{f}]} \frac{1}{L+K_{f}-1} ,&  k_f \in K_{f}\\
    \\
       \nu^{[l,k_{p}]} \frac{1}{K_{p,l}},     &  k_p\in K_{p,l}.
\end{cases}
\end{equation}
Therefore, the achievable rate of a generic user is given by 

\begin{equation}
\label{rate}
R^{[k]}_{\rm{VCSEL}} = \nu^{[k]} ~ \mathbb{E}\left[\log ~\det \left( \mathbf{I} + \mathrm{SNR} \right)\right].
\end{equation}
Notice that, the SNR of $ k_{f} $ and $ k_p $ users can be easily derived from  \eqref{fl} and \eqref{pl}, respectively.

\section{Problem Formulation}
We formulate an optimization problem that aims to maximize the  sum rate utility under the coverage of VCSELs. In this context, the utility function of a user rate  can be expressed as  
\begin{equation}
U^{[k]}_{\rm {VCSEL}} =  \varphi \left( \sum_{ l\in L} x^{[l,k]} \nu^{[l,k]} r^{[l,k]}_{\rm {VCSEL}}\right).
\end{equation}
where $ \varphi(.)= \log{(.)} $ is a logarithmic utility function that can achieve  proportional fairness among users \cite{879343}, and  $ x^{[l,k]} $ is defined as a variable that indicates whether VCSEL  $ l $ and user $ k $ are associated or not, i.e., $ x^{[l,k]} \in \{0,1\}$,  $ x^{[l,k]}=0 $ when user $ k $ is not assigned to VCSEL $ l $, otherwise, $ x^{[l,k]}=1 $. Moreover, $ r^{[l,k]}_{\rm {VCSEL}} $ is the achievable rate of user $ k $ that can be easily derived from \eqref{reseq} and \eqref{rate}. Therefore, the optimization problem is formulated as 
\begin{equation}
\label{pro-c}
\begin{aligned}
\max_{x,\nu} \quad  \sum_{k \in K}  \varphi \left( \sum_{ l\in L} x^{[l,k]} \nu^{[l,k]} r^{[l,k]}_{\rm {VCSEL}}\right) \\
 \textrm{s.t.}  \quad  \sum\limits_{l\in L}  x^{[l,k]}=1, ~~~~~~~~\forall k\in K_{P}, \\
   \sum_{k \in K_{p}} x^{[l,k]} \nu^{[l,k]} \leq 1, ~~~~~\forall l\in L, \\ 
 \quad x^{[l,k]}=1, ~~~~~~~~~~~~~~\forall k\in K_{f},  \\
 \sum_{k \in K_{f}} \nu^{[l,k]}\leq 1, ~~~~~~~~~~~~\forall l\in L, \\
 x^{[l,k]} \in \big\{0,1\big\}, 0 \leq \nu^{[l,k]} \leq 1, l \in L, k \in\{ K_{f}\cup K_{p}\}.\\
\end{aligned}
\end{equation}
The first constraint guarantees that each of the $ K_{p} $ users  connects only to one  VCSEL that maximizes the objective function, while the second constraint ensures that the total transmission probability for each VCSEL  $ l $ serving $ K_{p,l} $ users   is less than 1. Furthermore, the third constraint means that each of the $ K_f $ users  is connected to all VCSELs,  while the fourth constraint guarantees that the total transmission probability of the $ L $  VCSELs serving $ K_f $ users is less than 1. Finally, the last constraint describes the feasible region of the optimization variables.

The optimization problem in \eqref{pro-c} is a MINLP problem containing two binary variables $  x^{[l,k]} $ and $ \nu^{[l,k]} $ that rely on each other. In other words, the user assignment  has to be determined jointly with the resource allocation. This problem can be solved providing an optimal solution. However, it is such a complicated problem and not easily tractable where it involves high cost in term of complexity. Moreover, solving this problem means coordination among VCSELs is required, which is difficult to satisfy in  optical or RF cellular networks. In this following, the relaxation of the problem in \eqref{pro-c} is carried out through reformulating the main problem  into sub-problems that can be solved separately. 

\subsection{Decentralized resource allocation }
 \label{sec:Problem}
The Full decomposition method is adopted to solve the main problem \cite{6933944,9064520}. In the following the optimization problem is reformulated into two sub-problems:
\subsubsection{\textbf {Full connectivity users}}
The optimization problem in \eqref{pro-c} considering only the resource allocated to the $K_f  $ users  can be written as 
\begin{equation}
\label{pro-c1}
\begin{aligned}
\max_{\nu} \quad  \sum_{k_{f} \in K_{f}} \varphi \left( \sum_{ l\in L} \nu^{[l,k_{f}]} r^{[l,k_{f}]}_{\rm{VCSEL}}\right)~~~~ \\
\textrm{s.t.}  \sum_{k_{f} \in K_{f}} \nu^{[l,k_{f}]}\leq 1, ~~~~~~~~~~~~\forall l\in L, \\
 0 \leq \nu^{[l,k_{f}]} \leq 1, l \in L, k_{f} \in K_{f}.~~~
\end{aligned}
\end{equation}
The Lagrangian function for the optimization problem in \eqref{pro-c1} under its constraint can be written as
\begin{equation}
\label{pro-c2}
\begin{aligned}
f_{K_f}(\nu, \mu ) = \sum_{k_{f} \in K_{f}} \left[\varphi \left( \sum \limits_{ l\in L} \nu^{[l,k_{f}]} r^{[l,k_{f}]}_{\rm{VCSEL}}\right)-
  \sum_{ l\in L} \mu_l \nu^{[l,k_{f}]} \right]. 
\end{aligned}
\end{equation}
where $ \mu_l $ is 
the Lagrange multiplier or price associated with the constraint of the optimization problem in \eqref{pro-c1}. Then, the dual problem for the original problem in \eqref{pro-c1}  is given by 

\begin{equation}
\label{pro-c2}
\begin{aligned}
\min   g_{K_f}( \mu ) \\
 \mu \geq 0,
 \end{aligned}
\end{equation}
where $g_{K_f}( \mu )= \max f_{K_f}(\nu, \mu )$. Notice that, the problem in \eqref{pro-c2} is convex, which can find an upper bound on the optimal value. Therefore, the optimum  $ \nu^{*[l,k_{f}]} $ for a fixed value of the Lagrangian multiplier can be found through  using the Karush-Kuhn-Tucker (KKT) conditions solving the following  equation 

\begin{equation}
\frac{\partial \varphi \left( \sum_{ l\in L} \nu^{[l,k_{f}]} r^{[l,k_{f}]}_{\rm{VCSEL}}\right)}{\partial \nu }- \mu_l=0,
\end{equation}
where $  \nu^{[l,k_{f}]} \geq 0 $. After that, the optimum value of the multiplier $\mu_l  $ can be determined by  solving the problem in \eqref{pro-c2}. In this context, a gradient descent method can
be applied in order to update the multiplier $\mu  $ according to 

\begin{equation}
\mu_l(i+1)=\left[\mu_l(i) -\kappa_{\mu}(i)\left( 1- \sum_{k_{f}\in K_{f}} \nu^{*[l,k_{f}]} \right)\right]^{+},
\end{equation}
where $ i $ denotes is the iteration, $ \kappa_{\mu} $ is the step size and $ [.]^{+} $ is a projection on the positive orthant to account for the fact that $ \nu^{[l,k_{f}]} \geq 0 $.
\subsubsection{ \textbf{Partial connectivity users}}
For the users in the set $ K_p $, the optimization problem in \eqref{pro-c} can be written as 
\begin{equation}
\label{pro-c5}
\begin{aligned}
\max_{x,\nu} \quad  \sum_{k_{p} \in K_{p}}  \varphi \left( \sum_{ l\in L} x^{[l,k_{p}]} \nu^{[l,k_{p}]} r^{[l,k_{p}]}_{\rm{VCSEL}}\right) \\
 \textrm{s.t.}  \quad  \sum\limits_{l\in L}  x^{[l,k_{p}]}=1, ~~~~~~~~\forall k_{p}\in K_{P}, \\
   \sum_{k_{p} \in K_{p}} x^{[l,k_{p}]} \nu^{[l,k_{p}]} \leq 1, ~~~~~\forall l\in L, \\ 
 x^{[l,k_{p}]} \in \big\{0,1\big\}, 0 \leq \nu^{[l,k_{p}]} \leq 1, l \in L, k_{p} \in  K_{p}.\\
\end{aligned}
\end{equation}
To make this problem more tractable to solve, it is divided into two problems that can be solved separately on the user and VCSEL laser sides, respectively, as in the following:
\begin{itemize}
\item \textbf{User side:}
Let us consider $ K_{p,l} $ the number of users  associated with VCSEL $ l $. Assuming uniform resource allocation, the variable $ \nu^{[l,k_{p}]} $ is given by $ \frac{1}{K_{p,l}} $. Therefore, the Lagrangian function for the optimization problem on the user side is defined as
\begin{equation}
f_{K_p} (x,\eta)= \sum_{k_{p} \in K_{p}} \sum_{ l\in L} x^{[l,k_{p}]} \left(\varphi \left(  r^{[l,k_{p}]}_{\rm{VCSEL}}\right)- \eta_{l}\right),
\end{equation}
where  $ \eta_{l} $ is the Lagrange multiplier associated with the first constraint in the problem \eqref{pro-c5}. For solving the problem on the user side, each user $ k_{p}$, $ k_{p} \in K_{p}$ is associated with VCSEL $ l $ that satisfies the following

\begin{equation}
\label{pro-c66}
\begin{aligned}
\max_{x,\nu} \quad   \sum_{k_{p} \in K_{p}} \sum_{ l\in L} x^{[l,k_{p}]} \left(\varphi \left(  r^{[l,k_{p}]}_{\rm{VCSEL}}\right)- \eta_{l}\right) \\
 \textrm{s.t.}  \quad  \sum\limits_{l\in L}  x^{[l,k_{p}]}=1, ~~~~~~~~\forall k_{p}\in K_{P}, \\
 x^{[l,k_{p}]} \in \big\{0,1\big\}, l \in L, k_{p} \in  K_{p}.~~~~~~~\\
\end{aligned}
\end{equation}
Notice that, solving this problem, each VCSEL  might have a number  of the $ K_p $ users that request to be associated  with it at a given time. If there are multiple  VCSELs  that satisfy \eqref{pro-c66},  user $ k_p $ is associated with one of them due to the constraint $ \sum_{l\in L}  x^{[l,k_{p}]}=1 $. At this point, the users associated with VCSEL  $ l $ can be given by $ K_{p,l*}=  \sum_{k_{p} \in K_{p}} x^{[l*,k_{p}]} $. 

\item \textbf{VCSEL side:} Assuming the total number of the users calculated above forms the demand in term of the number of users. In this step, each VCSEL  calculates the supply number of users that can be served at a given time. Therefore, the optimization problem on the VCSEL  side can be formulated as  

\begin{equation}
\label{pro-c6}
\begin{aligned}
g_{K_p}(k_{p},\eta)=  \sum_{ l\in L} K_{p,l}  \left( \eta_{l}- \varphi (K_{p,l})\right).
\end{aligned}
\end{equation}
Therefore, the supply number of users that VCSEL $ l $ is capable of serving is obtained by solving
\begin{equation}
\label{pro-c6}
\begin{aligned}
\frac{\partial g_{K_p}(K_{p,l},\eta)}{\partial K_{p,l} }=0 \Rightarrow K_{p,l}=\exp(\eta_{l}-1).
\end{aligned}
\end{equation}
It is worth mentioning that the multiplier works as a bridge between the supply and demand numbers of users indicating the load on a VCSEL. For instance, $ K_{p,l}\leq K_{p,l*} $ means that VCSEL $ l $ is overloaded, and therefore, the multiplier goes up to increase the cost of using that VCSEL, otherwise, the multiplier goes down attracting more users to associate with VCSEL $ l $. As a result, the multiplier gets updated according to       
\begin{equation}
\label{pro-c6}
\begin{aligned}
\eta_{l}(i+1)= \left[\eta_{l}(i)-\kappa_{\eta}(i) \left( K_{p,l}-\sum_{k_{p} \in K_{p}} x^{[l*,k_{p}]} \right)\right]^{+},
\end{aligned}
\end{equation}
where $ \kappa_{\eta} $ is the step size.
\end{itemize}

\begin{table}
\centering
\caption{Simulation Parameters}
\begin{tabular}{|c|c|}
\hline
Parameter	& Value \\\hline
Bandwidth of VCSEL laser	& 5 GHz \\\hline
Wavelength of VCSEL laser	& 830 nm \\\hline
 VCSEL beam waist	& $ 5-30~ \mu $m \\\hline
Physical area of the photodiode	&15 $\text{mm}^2$ \\\hline
Receiver FOV	& 45 deg \\\hline
Detector responsivity 	& 0.53 A/W \\\hline
Gain of optical filter & 	1.0 \\\hline
Laser noise	& $-155~ dB/H$z \\\hline
\end{tabular}
\end{table}

\section{PERFORMANCE EVALUATIONS}
\label{sec:Pcom}
In this work a room with dimensions $ 5$m $ \times $ $5$m $ \times $ $3 $m  is considered, in which $ L=24 $ VCSELs  are distributed on the ceiling forming an array. These transmitters serve $ K $ users distributed randomly  on the receiving plane. Each user is equipped with a receiver composed of $ M=L $  infrared (IR) detectors, each with a distinct direction providing a channel response. Our simulation is carried out
for random independent snapshots of user distributions.     

\begin{figure}[t]
\begin{center}\hspace*{0cm}
\includegraphics[width=0.82\linewidth]{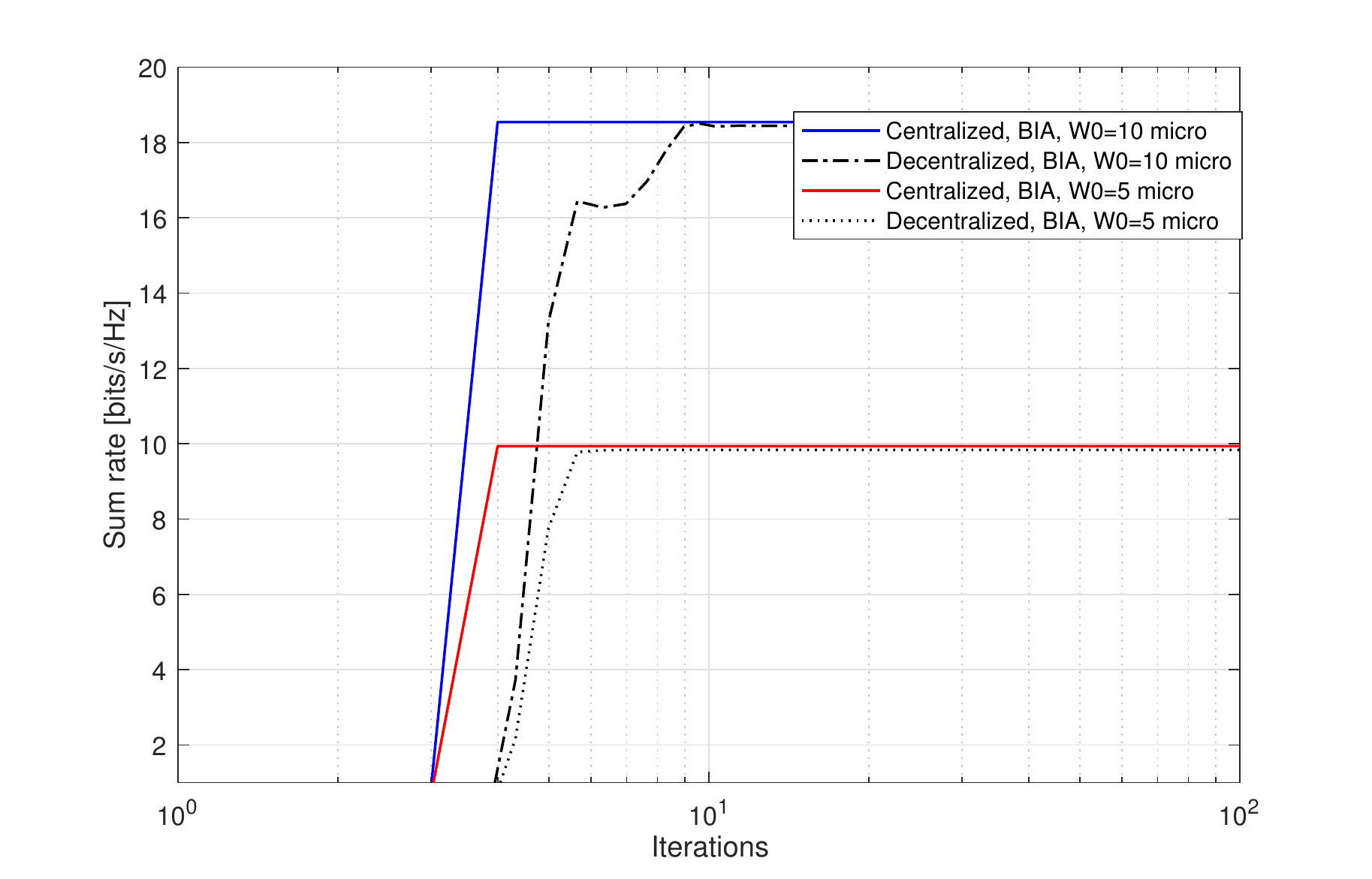}
\end{center}
\vspace{-2mm}
\caption{The sum rate of the VCSEL-based optical wireless network versus a set of iterations comparing centralized and decentralized algorithms. $ K=20 $.}\label{itra}
\vspace{-2mm}
\end{figure} 

In Fig. \ref{itra}, the achievable  sum rate is depicted against the number of iterations for the centralized and decentralized algorithms assuming two different values for the beam width of the VCSEL, i.e., $  W_{0}=\{5,10\}\mu $m. It is shown that the sum rate increases with $ W_{0} $ due to the fact that the narrow beam width of the VCSEL results in focusing more power towards the users, maximizing the received power of each user. Besides, the interference among multiple VCSELs decreases with the beam width where each VCSEL illuminates a more confined area. The figure further shows the convergence speed of the decentralized algorithm compared to the centralized algorithm where at iteration five, the solution provided by the decentralized algorithm is significantly close to the optimal solution regardless of the  beam width value. Therefore, the decentralized algorithm is chosen as a practical solution from now on due to its low cost in terms of complexity compared with the centralized solution.   

\begin{figure}[t]
\begin{center}\hspace*{0cm}
\includegraphics[width=0.82\linewidth]{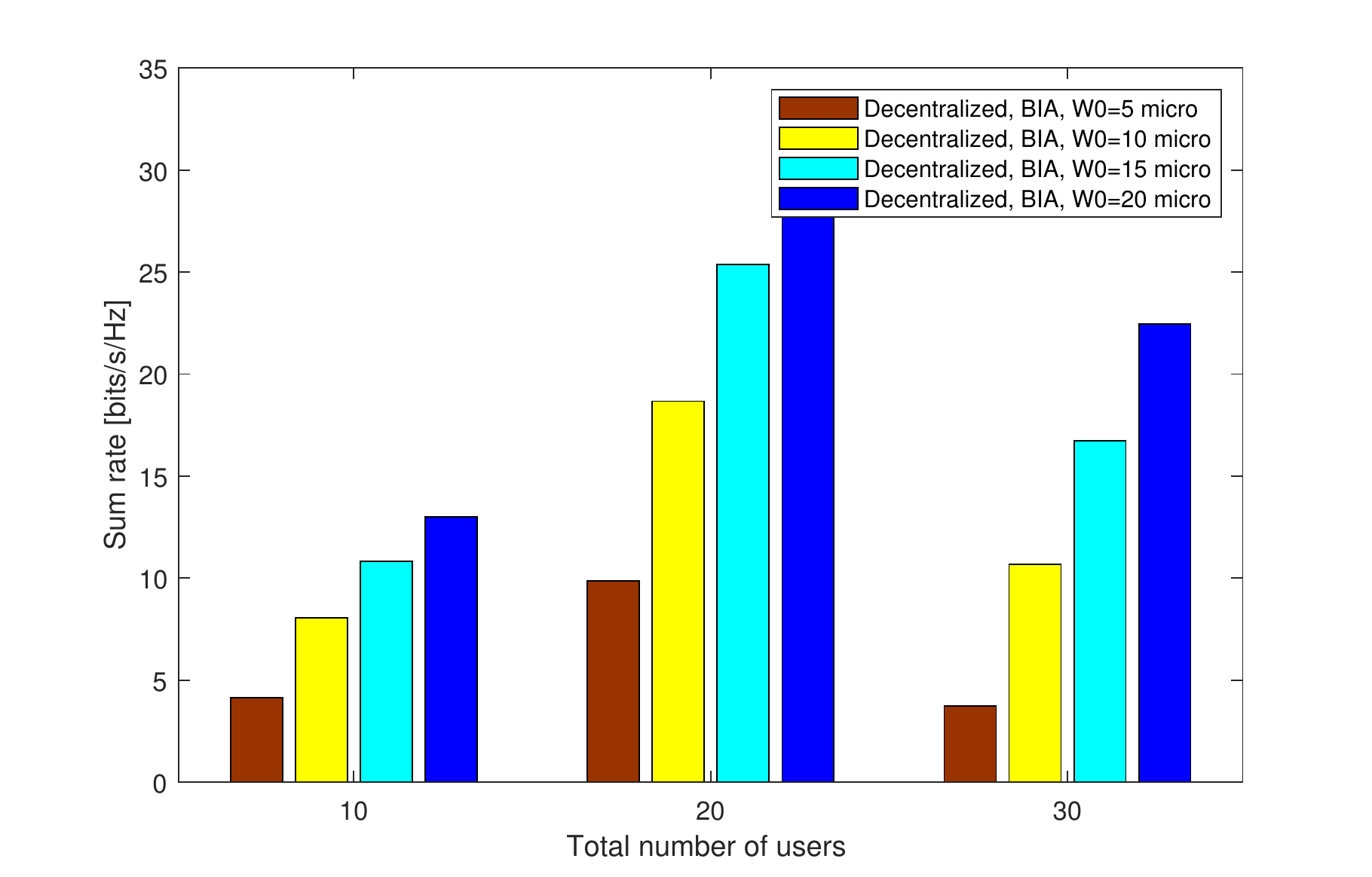}
\end{center}
\vspace{-2mm}
\caption{The sum rate of the VCSEL-based optical wireless network versus the number of users depicting the impact of the $ W_{0}$ value.}\label{userrat}
\vspace{-2mm}
\end{figure}

The achievable sum rate is also shown in Fig. \ref{userrat}, as the number of users in the network increases considering  four different values of the beam width, i.e., $  W_{0}=\{5,10,15,20\}\mu $m. It can be seen that when $ K=10  $ users, the achievable sum rate maximized using the decentralized algorithm  increases with the beam width due to the increase in the  power received by each user. Furthermore, when $ K=20 $ users, the achievable sum rate is higher  compared to  the scenario of the $ K=10$. This is because of  assigning the partial connectivity users based on maximizing their sum rate, in addition to maximizing the sum rate of the full connectivity users. In contrast, the sum rate decreases as the number of users increases to $ K=30 $, which is expected due to the implementation of BIA to align the interference among users. It is worth mentioning that the performance of BIA is limited in high density networks composed of  a high number of transmitters and users where BIA generates   a large supersymbol in order to serve all the users simultaneously, and therefore, the requirements of the channel coherence time become difficult to satisfy. Moreover, having a high number of users causes huge noise where each user might  subtract multiple interfering symbols \cite{GWJ11}.      

\begin{figure}[t]
\begin{center}\hspace*{0cm}
\includegraphics[width=0.82\linewidth]{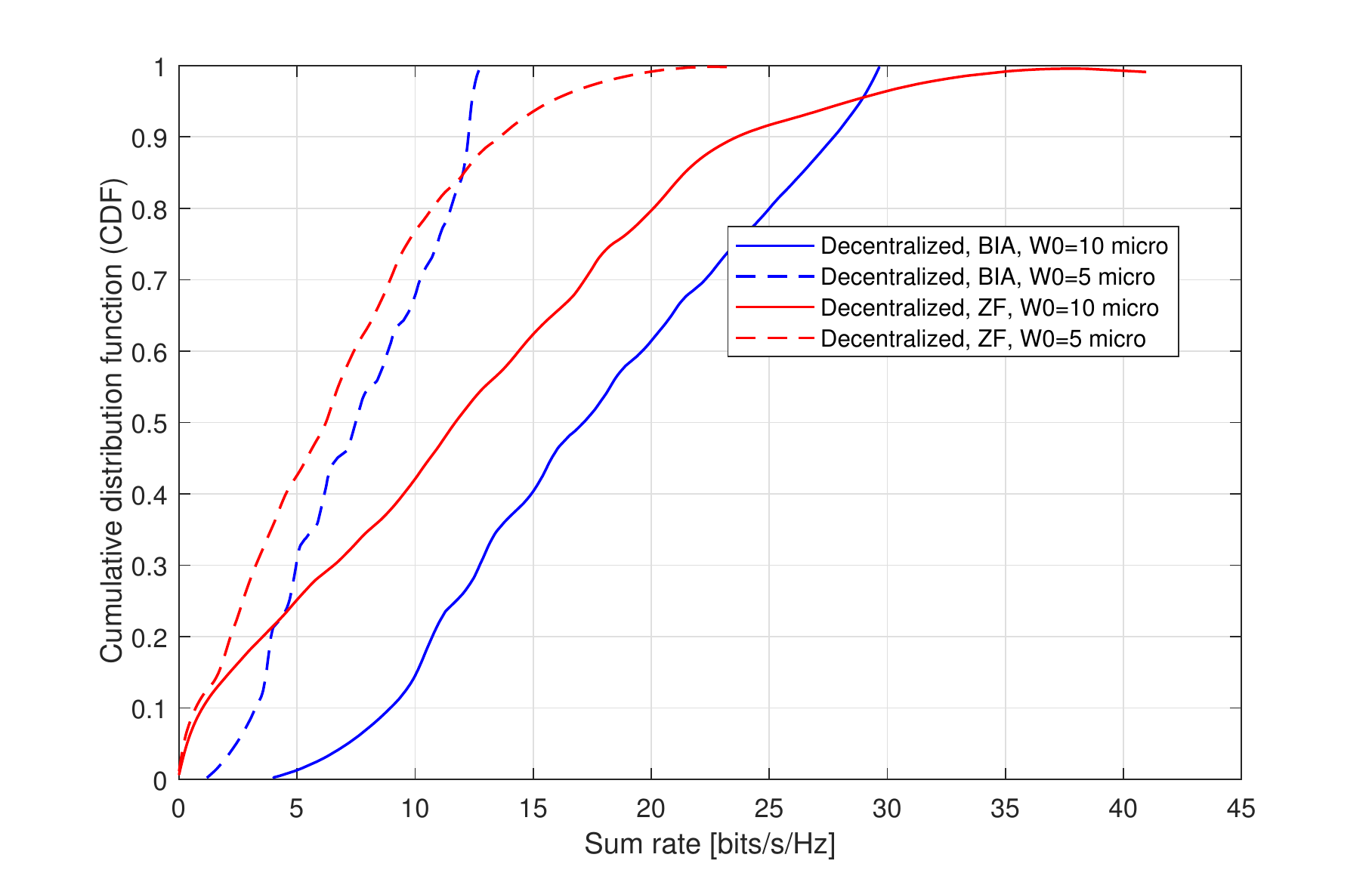}
\end{center}
\vspace{-2mm}
\caption{CDF of the sum rate in the VCSEL-based  wireless optical network. }\label{cdf}
\vspace{-2mm}
\end{figure}

Finally, the cumulative distribution function of the sum rate is depicted in Fig. \ref{cdf}, assuming $ K=20 $ users. It is shown that the performance of BIA is superior to ZF considering the same value of  the beam width due to the fact that the implementation of BIA achieves a fair rate distribution among the users. In addition, the performance of ZF is highly affected by the constraints of the optical channel such as the non-negativity of the transmitted signal and  the correlation among the channel responses of the users. Notice that BIA  naturally has a positive precoding matrix, and therefore,  its  performance  is less affected by the characteristics of the optical channel.   
\section{CONCLUSIONs}
\label{sec:CONCLUSION}
 In this paper, an optimization problem is formulated based on the connectivity of users in a laser-based optical network. First, the users are divided into two sets based on their connectivity. Then, an objective function representing the  utility-based sum rate maximization is derived. The formulated problem is MINLP, which is difficult  to solve. Therefore, a decentralized algorithm is proposed to solve the main problem providing a sub-optimal solution. The results demonstrate the convergence of the decentralized algorithm to the optimal solution provided by the centralized algorithm. Moreover, implementing BIA to align the interference among the users is more suitable than the ZF scheme.

\bibliographystyle{IEEEtran}
\bibliography{IEEEabrv,mybib}

\end{document}